\documentclass{ws-procs9x6}

\usepackage{latexsym}
\usepackage{amssymb}
\usepackage{amsmath}
\usepackage{epsfig}
\usepackage{cite}

\begin{document}


\title{The status of numerical relativity}


\author{Miguel Alcubierre}

\address{Instituto de Ciencias Nucleares, Universidad Nacional
Aut{\'o}noma de M{\'e}xico, A.P. 70-543, M{\'e}xico D.F. 04510, M{\'e}xico}


\maketitle

\abstracts{Numerical relativity has come a long way in the last three
decades and is now reaching a state of maturity.  We are gaining a
deeper understanding of the fundamental theoretical issues related to
the field, from the well posedness of the Cauchy problem, to better
gauge conditions, improved boundary treatment, and more realistic
initial data.  There has also been important work both in numerical
methods and software engineering.  All these developments have come
together to allow the construction of several advanced fully
three-dimensional codes capable of dealing with both matter and black
holes. In this manuscript I make a brief review the current status of
the field.}


\section{Introduction}

General relativity is a highly successful theory. It has radically
modified our vision of space and time and possesses an enormous
predictive power.  Its more important achievements include the
prediction of exotic objects such as black holes and neutron stars,
and the cosmological model of the Big Bang.  In addition, it predicts
the existence of gravitational waves, which will probably be detected
for the first time before the end of this decade.  Still, the field
equations of general relativity are extremely complex.  These
equations form a system of ten coupled non-linear partial differential
equations in four dimensions.  When fully expanded in a general
coordinate system they have thousands of terms.  Because of this,
exact solutions to these equations usually involve high degrees of
symmetry, either in space or in time: spherical or axially symmetry,
homogeneity, isotropy, staticity, stationarity, etc.  When one is
interested in studying highly dynamic systems that have little or
no symmetry it is simply impossible to solve the equations exactly and
one must fall back on numerical solutions.

Numerical relativity started in the 1960's with the pioneering work of
Hahn and Lindquist~\cite{Hahn64}, but it wasn't until the mid 1970's
that the first successful simulations were performed by Smarr and
Eppley in the context of the head-on collision of two black
holes~\cite{Eppley75,Smarr76}. Since then numerical relativity has
come a long way.  The availability of powerful super-computers,
together with an improved understanding of the fundamental theoretical
issues that underlie the field, have finally allowed simulations of
fully three-dimensional (3D) systems with strong and highly dynamical
gravitational fields. All this activity has arrived at the right time,
as the first generation of advanced gravitational wave detectors are
finally coming on line~\cite{LIGO_web,VIRGO_web,GEO_web,TAMA_web}.  We
are living a very exciting stage in the development of this field.

In this manuscript I will discuss the current status of numerical
relativity, starting from studies of theoretical issues having to do
with formulations of the equations, gauge conditions and initial data.
I will also discuss the latest achievements in the simulation of
astrophysical systems such as neutron stars and black holes.  Of
necessity this report will be very brief.  I will therefore limit
myself to discussing the main ideas and results, and will leave all
the details to the list of references at the end.


\section{The 3+1 decomposition}

The Einstein field equations are usually written in fully covariant
form, with no distinction between space and time.  This is elegant and
mathematically powerful, but it is not very useful when one is
interested in studying the evolution in time of a gravitational system
starting from some appropriate initial data, the so called ``initial
value problem''.

It is well known from the seminal work of
Choquet-Bruhat~\cite{Choquet52} that general relativity does allow an
initial value formulation.  Today there are three main procedures in
which one can obtain such a formulation: the ``Cauchy'' or ``3+1''
formalism, the ``conformal'' formalism, and the ``characteristic''
formalism. The 3+1 and conformal approaches share the property of
splitting spacetime into a foliation of spacelike hypersurfaces.  The
main difference lies in the fact that in the 3+1 approach these
hypersurfaces reach spatial infinity, while in the conformal approach
the hypersurfaces are hyperboloidal and reach future null infinity
instead. This allows the hypersurfaces to be conformally compactified,
permitting a rigorous treatment of fields at infinity and eliminating
the need for the boundary conditions at a finite distance that have to
be introduced in the standard 3+1 approach.  The characteristic
formalism, on the other hand, foliates spacetime using null cones
emanating from a central world tube and also allows a compactification
that brings null infinity to a finite coordinate distance.

Each approach has its strong points, but to date most work in
numerical relativity has been done using the 3+1 formalism, so here I
will concentrate on this approach.  However, interested readers are
referred to the excellent review of Winicour on the characteristic
approach~\cite{Winicour98} and the articles of Friedrich, Fraundiener
and Husa on the conformal
approach~\cite{Friedrich:2002xz,Frauendiener:2002iw,Husa:2002kk}.

In the 3+1 approach one introduces a global time function $t$ whose
levels sets are the hypersurfaces defining the foliation.  One then
defines three main ingredients: 1) The three-dimensional metric
$\gamma_{ij}$ ($i,j=1,2,3$) that measures distances within a given
hypersurface, 2) the ``lapse'' function $\alpha$ that measures proper
time between adjacent hypersurfaces, and 3) the ``shift vector''
$\beta^i$ that measures the relative speed between observers moving
along the normal direction to the hypersurfaces, and those keeping
constant spatial coordinates.  One then writes the four dimensional
metric as
\begin{equation}
ds^2 = \left( - \alpha^2 + \beta_i \beta^i \right) dt^2
+ 2 \, \beta_i \, dt dx^i + \gamma_{ij} \, dx^i dx^j \, ,
\end{equation}
where $\beta_i := \gamma_{ij} \, \beta^j$.

Having split the metric, one introduces the ``extrinsic curvature''
tensor $K_{ij}$ (also known as the ``second fundamental form'') that
measures how the spatial hypersurfaces are immersed in spacetime. This
tensor is given by the Lie derivative of $\gamma_{ij}$ along the time
lines, or in 3+1 language:
\begin{equation}
\partial_t \gamma_{ij} = - 2 \alpha K_{ij} + D_i \beta_j
+ D_j \beta_i \, ,
\label{eq:gdot}
\end{equation}
with $D_i$ the covariant derivative associated with $\gamma_{ij}$.

The next step is to decompose the Einstein equations.  Doing this one
finds that they naturally split in two groups.  One group involves no
time derivatives and represents constraints that must be satisfied at
all times.  The ``Hamiltonian constraint'' is given by (in units such
that $G=c=1$)
\begin{equation}
R + \left( \rm{tr} \, K \right)^2 - K_{ij} K^{ij} = 16 \pi \rho \, ,
\label{eq:ham}
\end{equation}
with $R$ the scalar curvature of the spatial geometry, $\rm{tr} \, K
\equiv \gamma^{ij} K_{ij}$ the trace of the extrinsic curvature and
$\rho$ the energy density of matter measured by the normal
observers. The ``momentum'' constraints take the form
\begin{equation}
D_j \left[ K^{ij} - \gamma^{ij} \rm{tr} \, K \right] =
8 \pi j^i \, ,
\label{eq:mom}
\end{equation}
with $j^i$ the momentum flux of matter measured by the normal
observers. The existence of the constraints implies, in particular,
that one is not free to specify the 12 dynamical quantities
$\{\gamma_{ij},K_{ij}\}$ as arbitrary initial conditions.  The
remaining 6 Einstein equations contain the dynamics of the system:
\begin{eqnarray}
\partial_t K_{ij} &=& \beta^a D_a K_{ij}
+ K_{ia} D_j \beta^a + K_{ja} D_i \beta^a - D_i D_j \alpha \nonumber \\
&+& \alpha \left[ R_{ij} - 2 K_{ia} K^a_j
+ K_{ij} \, \rm{tr} \, K \right] 
+ 4 \pi \alpha \left[ \gamma_{ij} \left( \rm{tr} \, S - \rho \right)
- 2 S_{ij} \right] \, ,
\label{eq:Kdot}
\end{eqnarray}
with $S_{ij}$ the stress tensor of matter.

It is important to mention that the Bianchi identities imply that the
evolution equations preserve the constraints, that is, if they are
satisfied initially they will remain satisfied at subsequent times.
The equations just described are know as the Arnowitt-Deser-Misner
equations, or ADM for short.  They represent the starting point of
practically all of 3+1 numerical relativity.  The reader interested in
seeing how these equations are derived is referred to the original ADM
article~\cite{Arnowitt62} or the classic review by York~\cite{York79}.


\section{Hyperbolic reductions}

The ADM evolution equations introduced in the previous section are in
fact highly non-unique.  This is because one can easily add arbitrary
multiples of the constraints to them, that is multiples of zero,
without affecting their physical solutions.  They key word here is
``physical'', since non-physical solutions, {\em i.e.}  those that do
not satisfy the constraints, will certainly change. Constraint
violating solutions, though not of interest physically, are
nevertheless inevitable in numerical simulations since truncation
errors imply that the constraints are never satisfied exactly.  Adding
constraints to the evolution equations can also seriously alter their
mathematical properties. This non-uniqueness of the evolution
equations is well known. For example, the original equations of
ADM~\cite{Arnowitt62} differ from those of York~\cite{York79}
precisely by the addition of a multiple of the Hamiltonian
constraint. The reformulation of York can be shown to be better
behaved mathematically~\cite{Frittelli97a} and has become the standard
form used in numerical relativity.

A key point that one has to worry about when studying the Cauchy
problem is the well-posedness of the system of evolution equations, by
which one understands that solutions exist (at least locally) and are
stable in the sense that small changes in the initial data produce
small changes in the solution. In this respect, one usually looks for
either symmetric or strongly hyperbolic systems of equations as such
systems are well posed under very general conditions~\cite{Kreiss89}.
In the case of general relativity, the problem of well-posedness was
studied already by Choquet-Bruhat in the 1950's~\cite{Choquet52}.  By
the mid 1980's a number of hyperbolic reductions were
known~\cite{Choquet62,Choquet80,Choquet83,Friedrich85}, though they
played a very minor role in numerical relativity (an excellent review
of the subject can be found in a recent article by Friedrich and
Randall~\cite{Friedrich:2000qv}).  In the early 1990's, Bona and Masso
started studying hyperbolic formulations for numerical
relativity~\cite{Bona89,Bona92,Bona94b}, and this effort was later
followed by a number of
people~\cite{Frittelli95,Abrahams96a,Friedrich96,Anderson99}. However,
a generalized interest of the numerical community in hyperbolicity had
to wait until 1999, when Baumgarte and Shapiro~\cite{Baumgarte99}
showed that a reformulation of the ADM equations proposed by Nakamura,
Oohara and Kojima~\cite{Nakamura87}, and Shibata and
Nakamura~\cite{Shibata95}, had far superior numerical stability
properties than ADM.  Baumgarte and Shapiro attributed this to the
fact that the new formulation, known as BSSN, had ``more hyperbolic
flavor''.  This was later put on firmer ground
in~\cite{Alcubierre99e,Sarbach02a,Beyer:2004sv}, and today it is
understood that ADM is only weakly hyperbolic (and thus not well
posed)~\cite{Frittelli:2000uj}, whereas BSSN is strongly
hyperbolic~\cite{Sarbach02a,Beyer:2004sv}.

The search for ever more general hyperbolic reductions of the Einstein
evolution equations has continued in the last few years, and today
many such formulations exist, several of which have dozens of free
parameters~\cite{Kidder01a,Sarbach02b}.  The focus has now changed to
finding hyperbolic formulations that allow dynamical gauge choices,
both for the lapse and the
shift~\cite{Lindblom:2003ad,Bona:2004yp,Alcubierre2003:hyperbolic-slicing},
and formulations that work well with fewer free parameters, maybe
keeping a second order in space
form~\cite{Nagy:2004td,Gundlach:2004jp}, or motivated by covariant
approaches like the recently developed ``Z4'' system of the Majorca
group~\cite{Bona:2003fj}.  At the same time, there is a growing
realization that well-posedness is not enough, as empirically some
hyperbolic formulations have proven to be far more robust than
others. Some work has been done on the analytic side trying to
understand what makes some hyperbolic formulations better suited for
numerical work.  In particular one can mention the work of Shinkai and
Yoneda~\cite{Shinkai02} and the work of Lindblom and
Scheel~\cite{Lindblom:2002et}.  Work in this area continues, and today
there is no consensus as to which, if any, is the best formulation of
the evolution equations for numerical purposes.


\section{Constrained evolution}

There is another approach, parallel but not inconsistent with the
drive for hyperbolic formulations.  This approach comes from the
observation that numerical instabilities are usually related with the
violation of the constraints.  It would then seem that imposing the
constraints at every time step should improve the stability of the
simulations.

Such ``constrained evolution'' is in fact a old approach, and is
commonly used in situations with high symmetry.  In fact, it is the
standard way to do numerical simulations in spherical symmetry, where
for specific gauge choices, {\em e.g} an areal radial coordinate, one
can solve for the metric directly from the Hamiltonian constraint and
ignore the evolution equations (in this case the Hamiltonian
constraint is an ordinary differential equation and therefore easy to
solve numerically).  In situations with less symmetry, and in
particular in fully 3D simulations, constrained evolution has not been
very popular since it involves solving systems of coupled elliptic
equations every time step, which makes it a hard problem.  The elliptic
equations can be solved, but it makes the evolution code extremely
slow (though this problem can be considerably ameliorated by using
different numerical techniques such as spectral methods).

Another reason why constrained evolution is less popular than ``free
evolution'' is that mathematical results on the well-posedness of
mixed elliptic-hyperbolic systems are sparse, and a priori there is no
good reason to expect that such constrained systems will be any better
behaved than standard approaches.  Still, a number of groups are
exploring the use of constrained evolution in axi-symmetry and even in
full
3D~\cite{Choptuik:2003as,Bonazzola:2003dm,Holst:2004wt,Matzner:2004uu},
and the idea seems to be gaining popularity.  It will be very
interesting to see if this approach can cure the stability problems
faced by free evolution codes.


\section{Gauge}

The Einstein equations provide us with evolution equations for the
spatial metric $\gamma_{ij}$ and extrinsic curvature $K_{ij}$.
However, they do not say anything about the evolution of the lapse
function $\alpha$ and shift vector $\beta^i$.  This is as it should
be, as the lapse and shift represent our freedom in choosing the
coordinate system, {\em i.e.} they are ``gauge'' functions.  Choosing
a good gauge is crucial when one solves the Einstein equations
numerically, and one must choose carefully if one wants to avoid
coordinate (and physical) singularities and to cover the interesting
regions of spacetime.  One can't just specify the lapse and shift as
{\em a priori} known functions of spacetime for a simple reason: Which
functions are a good choice?  The lapse and shift must therefore be
chosen dynamically as functions of the evolving geometry, that is, we
choose the coordinates as we go.

\subsection{Slicing conditions}

Since the seminal work of Smarr and York in
1978~\cite{Smarr78a,Smarr78b}, a lot has been learned about slicing
conditions.  Some classical choices such as maximal slicing are still
in use today.  Maximal slicing asks for the trace of the extrinsic
curvature to remain equal to zero, which implies that the normal
volume elements do not change. Through the ADM equations one finds
that this implies that the lapse function must satisfy the elliptic
equation:
\begin{equation}
D^2 \alpha = \alpha K_{ij} K^{ij} \; ,
\end{equation}
with $D^2$ the spatial Laplace operator.  Maximal slicing is an
extremely robust slicing condition, producing smooth evolutions
whenever in can be applied.  It also has the important property of
being ``singularity avoiding''.  This means that when a physical
singularity is approached, the slicing reacts by making the lapse
collapse to zero in that region, thus halting the evolution there
while allowing it to continue in regions away from the singularity. Of
course, there is a price to pay, and singularity avoiding slicings
develop a problem known as ``slice stretching'', where the radial
metric components grow rapidly as the slices are stretched between the
frozen inner regions and the evolving outer ones.  Maximal slicing
continues to be used to this day, and recently there has been some
important analytical insight into numerical results for black hole
evolutions using maximal slicing~\cite{Reimann:2003zd}.

The drive towards hyperbolicity and computational efficiency has made
hyperbolic slicing conditions increasingly popular in recent years,
and there have been a number of important developments in this
respect. In the first place, we have learned that it is much better to
prescribe the ``densitized lapse'' $\tilde{\alpha} := \alpha /
\det(\gamma)^{1/2}$ than the lapse itself, as a prescribed densitized
lapse allows hyperbolic formulations to be constructed while a
prescribed lapse does not.  Anderson and York have given important
insight as to why this is so~\cite{Anderson:1998we}, and Sperhake {\em
et al.} have recently shown dramatic numerical evidence of
this~\cite{Sperhake:2003fc}.  In the case of dynamical lapse
conditions, gauges of the Bona-Masso type~\cite{Bona94b}, $\partial
\alpha = - \alpha^2 f(\alpha) {\rm tr} K$, have also been shown to
allow hyperbolic formulations to be
constructed~\cite{Sarbach02b,Lindblom:2003ad,Alcubierre03c}.  These
type of gauge conditions are becoming very popular, in particular
``1+log'' slicing ($f=2/\alpha$), which has singularity avoiding
properties similar to maximal slicing but is far easier to implement
numerically.  Finally, there is now a better understanding of the
singularity avoidance properties of Bona-Masso type slicing
conditions, as well as the possibility of gauge shock
formation~\cite{Alcubierre02b} ({\em i.e.}  coordinate singularities
caused by the crossing of the characteristics associated with the
propagation of the gauge).

\subsection{Shift conditions}

Considerably less in known about shift conditions, and the classic
conditions of Smarr and York~\cite{Smarr78a} are still in use today in
one way or another.  In particular, the ``minimal distortion'' shift
condition, which was designed to minimize the distortion in the volume
elements over the hypersurface and requires that one solves a system
of three coupled elliptic equations, is still a common choice.
However, some recent results are important.

The main development has been the introduction of hyperbolic type
shift conditions associated with the BSSN formulation.  This
formulation introduces as auxiliary variables the contracted conformal
Christoffel symbols $\tilde\Gamma^i:=\gamma^{jk} \tilde
\Gamma^i_{jk}$, and shift conditions have been designed that make the
second time derivatives of the shift proportional to the time
derivatives of the $\tilde\Gamma^i$: $\partial^2_t \beta^i \sim
\partial_t \tilde\Gamma^i$.  It turns out that the principal part of
the operator on the right hand side of this equation is identical with
that of the minimal distortion condition, so these ``Gamma driver''
conditions can be seen as a hyperbolic version of minimal distortion.
These type of shift conditions are extremely easy to implement
numerically, in contrast with the difficulty involved in solving the
minimal distortion elliptic equations.  Moreover, they have proven to
be extremely robust in practice, controlling the slice stretching
associated with singularity avoiding slicings, and allowing very long
term evolutions of black hole spacetimes~\cite{Alcubierre02a}.
Because of this, such shift conditions are now being used by a large
fraction of existing 3D codes.  Recently, shift conditions of this
type have been shown to yield well-posed systems when considered
together with Einstein's evolution
equations~\cite{Lindblom:2003ad,Bona:2004yp}.

It has also become clear that 3D simulations of rotating and/or
orbiting compact objects require non-trivial shift vectors to avoid
large shears in the metric and allow long lived simulations.  The use
of co-rotating coordinates using a shift that approaches a rigid
rotation far away has been found to be crucial for achieving long
lived binary black hole
evolutions~\cite{Bruegmann:2003aw,Alcubierre2003:pre-ISCO-coalescence-times}.


\section{Boundaries}

Most physical systems one is interested in extend all the way to
infinity, but 3+1 simulations have a boundary at a finite distance
(however far). One must therefore impose some artificial boundary
condition at the edge of the computational domain.  This is not the
case in the conformal and characteristic approaches that allow one to
reach all the way out to null infinity, but it represents a serious
problem in standard 3+1 simulations.

For a very long time boundary conditions where not given the attention
the deserved by the numerical community.  The reason was that the
interior evolution was already extremely difficult and was usually
plagued with instabilities, so people considered themselves lucky if
the boundaries remained stable, however inconsistent they could be
with the physics.  With the development of more stable and robust
hyperbolic formulations for the interior evolution, the boundary
problem has come back to the attention of researchers in the field,
and today there is a lot of work in trying to ensure that the
boundaries are compatible with Einstein's equations (so-called
``constraint preserving'' boundaries).  The use of hyperbolic
formulations has also implied that one can separate incoming and
outgoing fields at the boundaries, allowing for boundary conditions
that are well-posed.  This means that there is now a rigorous way to
apply the intuitive notion of giving boundary data only for incoming
modes while allowing outgoing modes to leave the computational domain
undisturbed.

The crucial result in this field has been the first demonstration by
Friedrich and Nagy in 1999 of the well posedness of the
initial-boundary value problem for Einstein's
equations~\cite{Friedrich99}.  Since then a number of studies have
been made of maximally dissipative boundary conditions for linearized
gravity by Szilagyi {\em et al.}~\cite{Szilagyi00a} and Calabrese {\em
et al.}~\cite{Calabrese02c} and also non-linear gravity with harmonic
coordinates~\cite{Szilagyi02a,Szilagyi02b}. More recently other groups
have looked at the problem of boundaries, in particular one can
mention the work of Gundlach and Martin-Garcia in studying boundary
conditions for second order in space systems~\cite{Gundlach:2004jp},
and the work of Novak and Bonazzola on absorbing boundary conditions
for gravitational waves~\cite{Novak:2002hn}.


\section{Initial data}

Before starting a numerical simulation we need initial data that
represents the physical situation we want to evolve.  The existence of
the constraint equations implies that one needs to solve a system of
coupled elliptic equations to find such initial data.  Recently, there
have been important developments both in the general methods for
finding initial data, and in the search for better initial data for
black hole spacetimes.

\subsection{The thin sandwich formalism}

The classic method for finding initial data is the York-Lichnerowicz
conformal decomposition that finds initial data in the form of the
3-metric and the extrinsic curvature~\cite{Lichnerowicz44,York79}. In
this approach one gives as free data the conformal metric, the trace
of the extrinsic curvature and its transverse part, and solves the
constraints for the conformal factor and the longitudinal part of the
extrinsic curvature.

A very important recent development has been the introduction of the
``conformal thin sandwich'' approach by York, where initial data is
obtained in the form of the 3-metric and its time
derivative~\cite{York99}.  Pfeiffer and York have also shown the
equivalence of both conformal approaches~\cite{Pfeiffer:2002iy}.

\subsection{Binary black holes}

One of the most active areas of research in numerical relativity is
the simulation of the collision of orbiting black holes.  In the
following sections I will come back to this problem, but at this point
I will concentrate on the fact that in order to do those simulations
one must have some initial data.

Traditionally, initial data for black hole binaries has been obtained
using a topological construction.  This idea is based on the fact that
even a single Schwarzschild black hole contains a wormhole (the
``Einstein-Rosen'' bridge) joining two asymptotically flat universes.
Initial data for multiple black holes can then be constructed by
adding one new throat for each hole.  A formal procedure for doing
this is well known and comes in two ``flavors'': Misner type data
joins two universes by multiple wormholes, whereas Brill-Lindquist
type data joins one universe though a series of wormholes with
disjoint multiple other universes.  In the case of time-symmetric
initial data (corresponding to momentarily static black holes), both
types of data have an analytic expression. When the black holes have
momentum, as in the case of black holes in orbit around each other,
exact solution do not exist, but numerical solutions are not difficult
to find.  In 1994 Cook found Misner type data for black holes in
quasi-circular orbits using an effective potential
method~\cite{Cook94}.  Later, in 1997 Brandt and Bruegmann showed that
Brill-Lindquist type data could be obtained much more easily by
factoring out analytically the poles representing each black
hole~\cite{Brandt97b}, and in 2000 this ``puncture'' method was used
by Baumgarte to construct Brill-Lindquist type data for orbiting black
holes~\cite{Baumgarte00a}.  There has also been an effort to construct
initial data for binary black holes that does not use the topological
approach.  Matzner, Huq and Shoemaker have introduced the idea of
superposing single black hole data using Kerr-Schild type coordinates
and the solving the constraints~\cite{Matzner98a}.  This Kerr-Schild
type data has the advantage of reducing to stationary Kerr for
separated holes.

More recently, the thin sandwich approach has been used to construct
initial data for black holes that is more ``astrophysically
motivated''.  The main idea comes from the fact that black holes in
orbit have an approximate helical symmetry as they rotate. The
symmetry is not exact since the orbit shrinks as the system emits
gravitational waves, but for sufficiently separated holes this happens
very slowly.  One can then exploit the fact that an approximate
helical Killing field exists to obtain black hole initial data in a
co-rotating frame. Grandclement, Gourgoulhon and Bonazzola have used
this approach to construct binary black hole initial data of the
Misner type~\cite{Grandclement02}. This data seems to agree closer
with high order post-Newtonian predictions than the data based on the
effective potential method.  Tichy and Bruegmann have also applied
this approach to puncture type data~\cite{Tichy:2003qi}, and Yo {\em
et al.} have done the same for Kerr-Schild type data~\cite{Yo:2004ng}.


\section{Horizons}

When studying the numerical simulations of black hole spacetimes it is
important to locate the horizons associated with them.  Traditionally,
the apparent horizon has been used as an indicator of the presence of
a black holes since such horizons are defined locally and can be found
on a given hypersurface during the evolution. On the other hand, the
existence of an event horizon represents the true definition of a
black hole, so locating the event horizon would be much better.
However, the global nature of the definition of an event horizon means
that one can not locate them during the evolution since one needs to
know the full history of the spacetime, and that is precisely what one
is trying to solve for.  Still, algorithms exist to find event
horizons {\em a posteriori}.  The basic idea was developed by Libson
{\em et al.}~\cite{Libson94a}, and consists on evolving a null surface
backwards in time once an evolution is finished.  As the event horizon
is an attractor for light rays moving back in time, such null surfaces
rapidly converge to the event horizon and track it all the way to its
initial formation (or to the initial data if already present there).
An important recent event has been the development of a 3D event
horizon finder using these ideas by Diener~\cite{Diener03a}.

Another development has been the introduction of the theoretical
framework of isolated and dynamic horizons of Ashtekar {\em et
al.}~\cite{Ashtekar98a,Ashtekar02a} to numerical relativity by Dreyer
{\em et al.}~\cite{Dreyer02a}.  The formalism permits to calculate
both the mass and the angular momentum associated with horizons found
during a numerical simulation. These techniques have also recently
been applied to the binary coalescence of orbiting black
holes~\cite{Alcubierre2003:pre-ISCO-coalescence-times}.


\section{Evolving black hole spacetimes}

The simulation of black holes spacetimes has been a major theme in
numerical relativity since its beginnings.  The pioneering work of
Hahn and Lindquist in the 1960's~\cite{Hahn64}, as well as the work of
Smarr and Eppley in the 1970's~\cite{Eppley75,Smarr76}, was precisely
on simulations of collisions of black holes.

Black hole spacetimes are in vacuum, so the simulation of these
systems has the advantage of not having to deal with complicated
hydrodynamics.  On the other hand, black holes have a singularity
where the gravitational field becomes infinite, and dealing with this
singularity is far from trivial.  The traditional technique to evolve
black holes has been the use of singularity avoiding slicing
conditions such as maximal slicing, but this has the weakness of
leading to slice stretching which eventually becomes a serious problem
on its own.  An alternative approach is known as ``singularity
excision'' (previously called ``apparent horizon boundary
condition'').  The original idea has been attributed to Unruh by
Thornburg~\cite{Thornburg87,Thornburg93}, and consists on placing a
boundary inside the black hole, thus excising its interior from the
computational domain, and also using a non-zero shift vector that
keeps the horizon roughly in the same coordinate location during the
evolution.  Since no information can leave the interior of the black
hole, excision should have no effect on the physics outside. Black
hole excision was first attempted successfully by Seidel and Suen in
spherical symmetry~\cite{Seidel92a}, and was later studied in more
detail by Anninos et.al.~\cite{Anninos94e}.  Work in singularity
excision for 3D codes continues to this day.

In the early 1990's, a lot of work was done in simulating single
static or perturbed black holes~\cite{Anninos93c}, as well as more
advanced versions of the head-on
collision~\cite{Anninos93b,Anninos94b}.  By the middle of that decade,
the simulation of the inspiral collision of orbiting black holes was
identified as a key project, motivated by the fact that the large
interferometric gravitational wave detectors were finally starting
construction.  This led to the NSF ``Binary Black Hole Grand
Challenge'' project, that brought together a large number of
institutions in the United States. The problem turned out to be
considerably more complicated than at first thought, among other
reasons because the issue of the well-posedness of the evolution
equations was not yet fully understood.  Still, a number of important
developments did take place, among which one can mention improved
singularity excision techniques~\cite{Cook97a}, and a long term stable
evolution of a single black hole using the characteristic
formalism~\cite{Gomez98a}.
 
Since the end of the Grand Challenge project, a number of groups have
continued independently trying to simulate the inspiral collision of
two black holes.  In the last few years there has been rapid progress,
coming on the one hand from the use of more robust well-posed
formulations such as BSSN~\cite{Baumgarte99}, a better understanding
of black hole excision
techniques~\cite{Alcubierre00a,Alcubierre01a,Calabrese:2003a,Thornburg2004:multipatch-BH-excision,Alcubierre2003:BBH0-excision},
and the development of improved gauge conditions, particularly related
to the choice of the shift vector~\cite{Alcubierre02a}.  This led to
the first simulations of the grazing collision of two black holes done
in 2000 by Brandt {\em et al.}~\cite{Brandt00} and Alcubierre {\em et
al.}~\cite{Alcubierre00b}.

\begin{figure}
\epsfxsize=100mm
\epsfysize=60mm
\centerline{\epsfbox{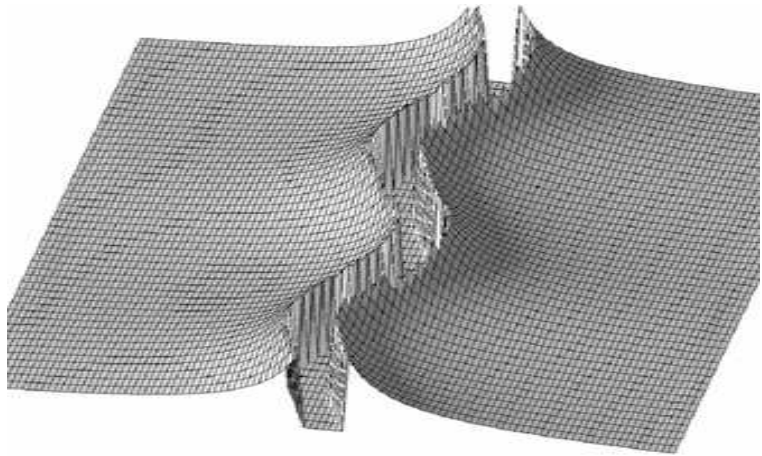}}
\caption{Space-time plot of the BSSN conformal factor $\phi$ for a
single black hole oscillating in coordinate space. (Figure from
Ref.~\protect \cite{Shoemaker2003a})}
\label{fig:wobbling}
\end{figure}

\begin{figure}
\epsfxsize=100mm
\epsfysize=50mm
\centerline{\epsfbox{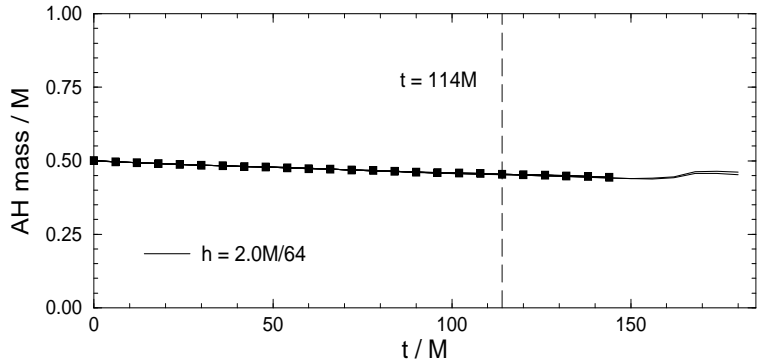}}
\caption{Evolution of the AH mass for one of the individual black
holes in an orbiting binary. The evolution lasts longer than one
orbital period of $114M$ defined by the initial data, and no common
horizon is found (Figure from Ref.~\protect \cite{Bruegmann:2003aw})}
\label{fig:orbit}        
\end{figure}

\begin{figure}
\epsfxsize=100mm
\epsfysize=60mm
\centerline{\epsfbox{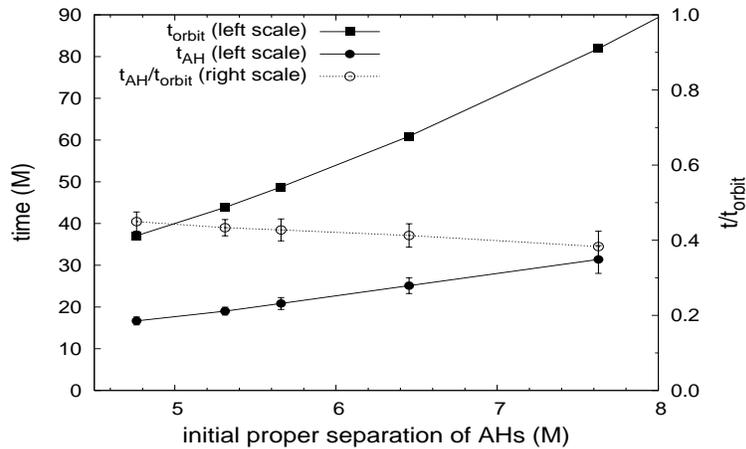}}
\caption{The time to appearance of a common apparent horizon for black
hole in quasi-circular orbits for progressive separations, starting
from the Cook/Baumgarte innermost stable circular orbit. Filled
circles indicate the results of numerical simulations. The upper line
indicates the expected orbital period, based on the initial angular
velocity.  Empty circles indicate the fraction of an orbit before
common horizon formation. As the black holes are further separated,
they take longer to coalesce, but the do it in a smaller fraction of
an orbit (Figure from Ref.~\protect
\cite{Alcubierre2003:pre-ISCO-coalescence-times})}
\label{fig:coalescence}
\end{figure}

Research has now moved into the realm of orbiting black holes, with
two different approaches taking the lead, both of them based on the
BSSN formulation.  One approach uses advanced excision techniques to
move black holes across the computational domain and has recently
achieved a true milestone by having single black holes move in
oscillating and even circular trajectories (controlled by gauge
choices) in a stable way for essentially unlimited
time~\cite{Shoemaker2003a} (See Fig.~\ref{fig:wobbling}).  A second
approach uses less advanced excision techniques and attempts to keep
the orbiting black holes on a fixed coordinate location by using
instead advanced gauge choices and a co-rotating frame.  This approach
is further along the road, and has finally led to the first ever
simulation of a full orbit of two black holes by Bruegmann {\em et
al.}~\cite{Bruegmann:2003aw} (See Fig.~\ref{fig:orbit}).  More
recently, using the same approach but an independent code, Alcubierre
{\em et al.} have studied sequences of black holes in quasi-circular
orbits of the Cook/Baumgarte type and have come to the surprising
discovery that the black holes are in fact not orbiting, but are
instead in an extended
plunge~\cite{Alcubierre2003:pre-ISCO-coalescence-times} (See
Fig.~\ref{fig:coalescence}) .  Other groups are also working on the
binary black hole problem using a variety of techniques, among them
one can mention the NASA-Goddard group that uses a BSSN code with mesh
refinement~\cite{Imbiriba:2004tp}, and the Cornell/Caltech
collaboration that uses advanced pseudo-spectral methods together with
a multi-parameter hyperbolic formulation of the evolution
equations~\cite{Scheel2002a}.

\begin{figure}
\epsfxsize=90mm
\epsfysize=60mm
\centerline{\epsfbox{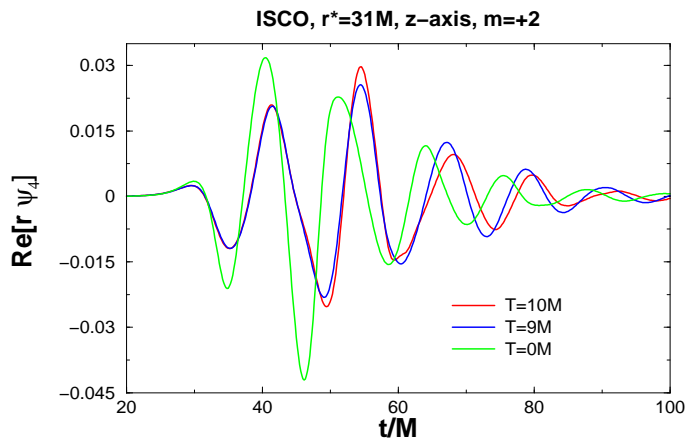}}
\caption{Wave forms for the Cook/Baumgarte innermost stable circular
orbit using the Lazarus technique.  The different lines show the
effects of extracting data for a perturbative evolution at different
times. (Figure from Ref.~\protect \cite{Baker:2002qf})}
\label{fig:lazarus}
\end{figure}

A very important development has been the combination of fully
nonlinear techniques with perturbation techniques by Baker {\em et
al.} to study the black hole coalescence
problem~\cite{Baker00b,Baker:2001nu,Baker:2001sf,Baker:2002qf}.  This
approach evolves the black holes using a full non-linear code only to
a ``linearization'' time where the two holes are close enough that the
system can be treated as a single distorted Kerr black hole, after
which a perturbative approach is used to complete the evolution.  The
approach has been called ``Lazarus'', making reference to the fact
that a ``dying'' non-linear evolution (about to crash because of
instabilities) can be resurrected by continuing the evolution using
perturbation methods.  The Lazarus approach has already given the
first complete gravitational wave-forms to come from the coalescence
of two orbiting black holes (See Fig.~\ref{fig:lazarus}).

\section{Relativistic hydrodynamics}

Apart from black hole spacetimes, there is also a lot of interest in
systems with matter.  In particular, there has been considerable work
on fully 3D general relativistic hydro-dynamical codes.  At this time,
several codes exist that are capable of simulating core collapse,
rotating stars and binary star mergers.  Among the groups with
advanced 3D hydro codes one can mention the Tokyo
group~\cite{Shibata:2003iy,Shibata:2003iw,Shibata:2003ga}, the
Illinois group~\cite{Duez04,Duez04hydro}, the EU-Network
group~\cite{Baiotti04} and the JPL-WashU group~\cite{Miller:2003vc}.

It is interesting to notice the convergence of the techniques used in
the different codes: All the codes mentioned above use some variety of
the BSSN evolution system for the evolution of the geometry; all use
either maximal slicing or a hyperbolic slicing condition of the
Bona-Masso type (1+log for the most part); all use a non-zero shift
vector, with the shift controlled by an evolution equation of the
Gamma-driver type; advanced numerical techniques of the type of
high-resolution shock-capturing schemes are becoming standard for the
hydro-dynamical equations; several codes have already, or are
implementing, fixed mesh refinement techniques; several codes have
excision techniques for handling black hole formation.

\begin{figure}
\epsfxsize=90mm
\epsfysize=70mm
\vspace{8mm}
\centerline{\epsfbox{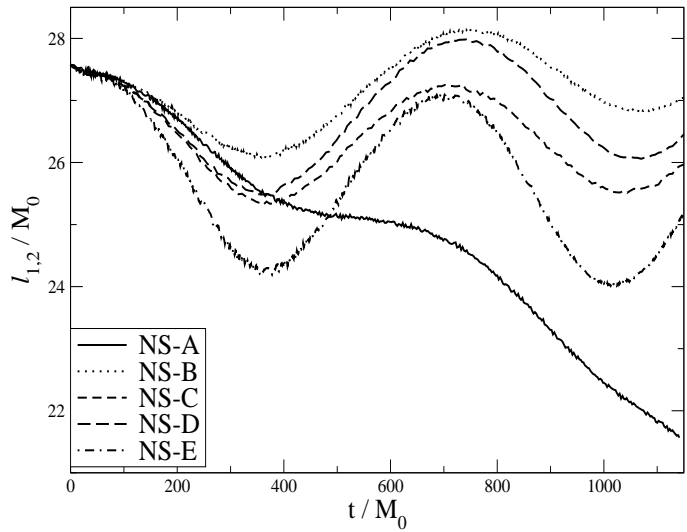}}
\vspace{2mm}
\caption{Evolution of geodesic distance between the centers of two
orbiting neutron stars as a function of time, for various initial data
sets representing the same physical situation but differing in grid
resolution and position of outer boundary.  Model NS-A shows a
qualitatively very different behavior, attributed to the fact that in
this case the boundaries of the computational domain were much further
out than in the other cases. (Figure from Ref.~\protect
\cite{Miller:2003vc})}
\label{fig:neutron}
\end{figure}

Among the most important results so far one can mention the studies of
collapse of super-massive stars~\cite{Duez04hydro} and the studies of
collapsing rapidly rotating neutron stars~\cite{Baiotti04}. A
particularly interesting result has been the study of the strong
effects of boundary conditions on the orbital dynamics of binary
neutron stars by Miller {\em et al.}~\cite{Miller:2003vc} (see
Fig.~\ref{fig:neutron}).  The field is moving rapidly, and many
interesting results are sure to follow in the near future.

\section{Conclusions}

Numerical relativity has come a long way in the last three decades.
We are gaining a deeper understanding of the fundamental theoretical
issues that underlie the field, such as well posedness of the
evolution equations, gauge conditions, boundary conditions and initial
data.  At the same time, several advanced 3D codes exist that can
handle both black holes and hydrodynamics.  These codes are capable of
singularity excision and mesh refinement, and have complex analysis
tools incorporated: apparent and event horizon finding, horizon
studies (mass, angular momentum), gravitational wave extraction, etc.
The field is now reaching a state of maturity.  In the next few years
we hope to be able to simulate interesting astrophysical systems and
extract gravitational waveforms that can be compared with the
observations coming from the large interferometric detectors.
Numerical relativity is truly entering a golden age!


\section*{Acknowledgments}

It is a great pleasure to thank the organizers of the GR17 conference
for their invitation to give a talk and write these notes.  I also
wish to thank Pablo Laguna, Bernd Bruegmann, Carlos Lousto, Manuela
Campanelli, John Baker and Mark Miller for allowing me to show plots
from their work in this manuscript. Finally, I want to thank the whole
numerical relativity community for their enthusiastic response to my
request for information about their current work. This work was
supported in part by DGAPA-UNAM through grants IN112401 and IN122002.


\bibliographystyle{bibtex/prsty}
\bibliography{bibtex/references}


\end{document}